# A Light-Emitting-Diodes-Integrated Silicon Carbide Insulated Gate Bipolar Transistor

Guoliang Zhang[1], Zhanwei Shen[2], Yujian Chen[1], Yufeng Qiu[1], Feng Zhang[1,3*], and Rong Zhang[1*]

[1] College of Physical Science and Technology, Xiamen University, Xiamen, 361005, People's Republic of China
[2] Laboratory of Semiconductor Material Sciences, Institude of Semiconductors, Chinese Academy of Sciences, Beijing, 10083, People's Republic of China
[3] Jiujiang Research Institute, Xiamen University, Jiujiang, 332000, People's Republic of China

E-mail: zwshen@semi.ac.cn; fzhang@xmu.edu.cn; rzhangxmu@xmu.edu.cn

**Abstract**

A light-emitting-diodes (LEDs)-integrated silicon carbide (SiC) insulated gate bipolar transistors (LI-IGBT) is proposed in this paper. The novelty of the LI-IGBT depends on the photogeneration effect of III-nitride LEDs embedded in the poly-Si regions of IGBT. Then, the photogenerated carriers are formed in the JFET region and the drift layer, indicating the increase of the conductivity in LI-IGBT as compared with the SiC IGBT with hole-barrier layer (H-IGBT) and the SiC IGBT with charge storage layer (CSL-IGBT). The static simulation results show that the electron density of the LI-IGBT at the middle of the drift layer is separately 17.44 times and 15.81 times higher than those of the H-IGBT and CSL-IGBT, yielding 40.91% and 37.38% reduction of forward voltage drop, respectively, and also, the LI-IGBT shows 304.59% and 263.67% improvements in BFOM as compared with CSL-IGBT and H-IGBT, respectively. For the dynamic simulation in one cycle, the loss of LI-IGBT is separately reduced by 6.57% and 8.57% compared to H-IGBT and CSL-IGBT. Meanwhile, the relationship between $V_{C(sat)}$ and $E_{turn-off}$ can be optimized by adjusting collector doping and minority carrier lifetime. These results reveal that the proposed SiC IGBT will be more suitable for ultra-high voltage application.

Keywords: Silicon carbide, Insulated gate bipolar transistors (IGBTs), Light-emitting diodes (LEDs)

## 1. Introduction

Due to superior material properties, such as high critical electric field and high thermal conductivity, silicon carbide (SiC) has been demonstrated to be a promising material for next generation power semiconductor applications[1]. At present, for low and medium voltage applications, SiC metal-oxide-semiconductor field effect transistor (MOSFET) is commercially available and increasingly mature in new energy vehicles and photovoltaic systems[2-6]. However, for high and ultrahigh voltage power applications, SiC unipolar devices feature thick drift layers with low doping concentrations leading to high drift layer resistance[7-9]. Thus, reducing ON-state loss in the power system is seriously hindered[10]. By introducing conductivity modulation into the thick drift layer and utilizing reliable gate processing in the MOS technique, SiC insulated gate bipolar transistor (IGBT) offers excellent benefits in the improvement of the ON-state characteristic and high robustness for high and ultrahigh voltage as compared to MOSFET with the same rated blocking voltage[11].

It has been demonstrated in silicon that IGBT with the current spreading layer (CSL) or super junction (SJ) structure enables the device to have low forward voltage due to low

JFET resistance or reduced drift layer resistance, respectively. Therefore, these conceptual prototypes can be imitated in SiC for the same purpose of lowering ON-state loss[12-15]. However, because of the wide band gap property of silicon carbide and its low intrinsic carrier density, it is difficult to generate high-density carriers by thermal excitation. Furthermore, it is not easy to generate the conductivity modulation in the drift layer of the conventional IGBT, considering the fatal flaw of the thick epitaxial layer with a low carrier lifetime in SiC[16]. These results reveal that high drift resistance in SiC is not identically reduced by the approaches as expected in silicon.

This work proposes an ultraviolet light-emitting-diodes (LEDs)-integrated SiC IGBT structure (LI-IGBT). The integration process of LED is compatible with SiC power device manufacturing. Because III-nitride LEDs for Micro-LEDs can produce efficient emissive cells typically grown on the AlN buffer layer on SiC[17, 18]. Meanwhile, SiC IGBT has a large cell pitch of above 10 μm and the SiC is sensitive to ultraviolet light. In this regard, the on-resistance of SiC IGBT can be reduced by amounts of the photogenerated carriers generated under the ultraviolet light illumination and those carriers can promote the conductivity modulation in the drift layer.

## 2. Device structure and mechanism

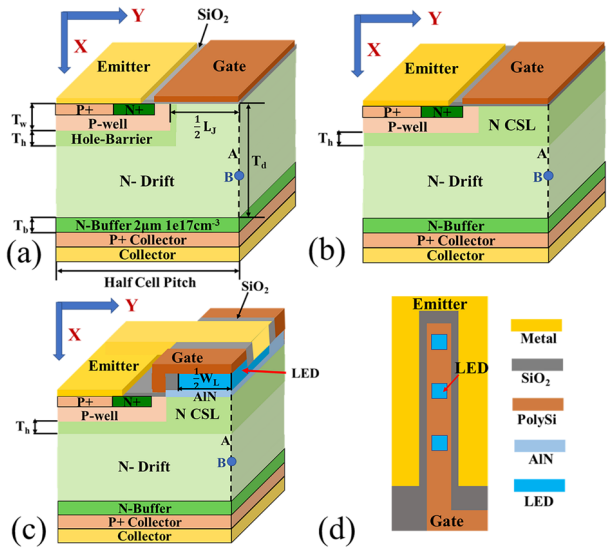

**Figure. 1.** Structures of (a) the LI-IGBT, (b) the top view of LI-IGBT and (c) the H-IGBT. And (d) the CSL-IGBT.

Schematic structures of SiC LI-IGBT, the SiC IGBT with hole-barrier layer (H-IGBT)[19], and the SiC IGBT with charge storage layer (CSL-IGBT)[20] are separately illustrated in figures. 1(a),(c),(d). The top view of LI-IGBT is illustrated in figure. 1(b). In this work, the n-type doping of $2\times10^{14}$ cm$^{-3}$ and thickness of 200-μm 4H-SiC epi-layer is selected to sustain the 20-kV blocking voltage. The gate oxide thickness and the width of channel are fixed to 50 nm and 0.5 μm, respectively. To attain superior tradeoff between the ON-state and OFF-state performances, an n-type charge storage layer (CSL) with the doping of $1\times10^{16}$ cm$^{-3}$ as a hole barrier is implemented.

Meanwhile, a 100-nm thick AlN is severed as a buffer layer between the 4H-SiC epi-layer and LEDs. The anode and cathode of LEDs are connected with the gate and emitter of LI-IGBT, respectively. The active width of the LED is 10 μm, which ensures that the LED can be integrated above the 4H-SiC and the channel of IGBT can work normally. In order to guarantee all the drift layer can be covered by the ultraviolet light, the orthogonal spacing of the LED is set to 500 μm. The main device parameters are shown in Table 1.

**Table 1.** Major device parameters

| Parameters | LI-IGBT | H-IGBT | CSL-IGBT |
|---|---|---|---|
| Cell Pitch (μm) | 25 | 25 | 25 |
| $L_j$ (The length of JFET region) (μm) | 10 | 10 | 10 |
| $W_{LED}$ (The width of LEDs) (μm) | 10 | - | - |
| $T_W$ (The thickness of well) (μm) | 0.6 | 0.6 | 0.6 |
| $T_h$ (The thickness of hole-barrier layer or charge storage layer) (μm) | 0.5 | 0.5 | 0.5 |
| $T_d$ (The thickness of drift layer) (μm) | 200 | 200 | 200 |
| $T_{bu}$ (The thickness of buffer layer) (μm) | 2 | 2 | 2 |
| The length of channel (μm) | 0.5 | 0.5 | 0.5 |
| The thickness of oxide (μm) | 0.05 | 0.05 | 0.05 |
| The thickness of AlN (μm) | 0.1 | - | - |
| Doping of charge storage layer/ hole-barrier layer (cm$^{-3}$) | $1\times10^{16}$ | $1\times10^{16}$ | $1\times10^{16}$ |
| Doping of n- drift layer (cm$^{-3}$) | $2\times10^{14}$ | $2\times10^{14}$ | $2\times10^{14}$ |
| Doping of n-buffer layer (cm$^{-3}$) | $1\times10^{17}$ | $1\times10^{17}$ | $1\times10^{17}$ |
| Doping of collector layer (cm$^{-3}$) | $1\times10^{19}$ | $1\times10^{19}$ | $1\times10^{19}$ |
| Doping of P well (cm$^{-3}$) | $1\times10^{17}$ | $1\times10^{17}$ | $1\times10^{17}$ |

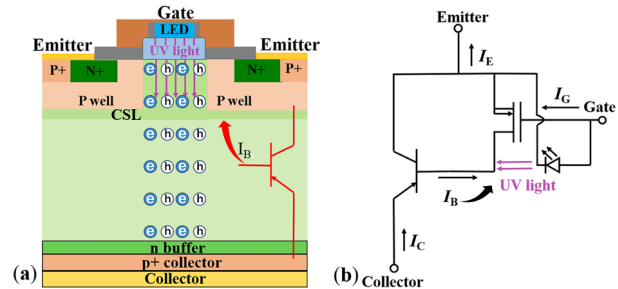

**Figure. 2.** (a) The ON-state mechanism of LI-IGBT. (b) Equivalent circuit at ON-state of LI-IGBT.

The physical structure and conduction mechanism of LI-IGBT are depicted in figure. 2(a). In ON-state, the IGBT and LED are simultaneously driven by applying the positive voltage on the gate or the anode terminals of the devices. Then, the emitted ultraviolet light can be vertically irradiated into the whole drift regions of the LI-IGBT, resulting in amounts of photogenerated carriers. To clarify the light loss effect inside the device, we also considered the absorption of the light by the AlN nucleation layer and the reflection of the light at the interfaces[21-23]. Figure. 2(b) illustrates equivalent circuit at ON-state of LI-IGBT. The integrated LED is inserted between the gate and emitter terminals. The photogenerated carriers are stimulated into the JFET regions and the drift regions by the ultraviolet light, increasing current ($I_B$) in the base region of the parasitic PNP bipolar junction transistor (BJT). The increased current improves the conduction characteristic in the thick drift layer. Thus, the forward voltage drop of the IGBT is reduced.

## 3. Simulation results and discussion

The electrical characteristics of SiC LI-IGBT, H-IGBT, and CSL-IGBT are simulated and compared using the numerical simulation software via TCAD[24]. Shockley-Read-Hall, Auger, and incomplete dopant ionization models are enabled in all structures. The specified monochromatic source is activated in the optical generation model. Then the quantum yield model takes the band gap into account so as to describe the number of absorbed photons converted to generated electron-hole pairs. The optical power density of the integrated LED is calculated by utilizing the effective area and the optical power of the LED. The threshold voltage is 6 V for the integrated LEDs in the wavelength band range from 280 nm to 320 nm[25, 26] and is 4 V for those in the wavelength band above 320 nm[27]. Meanwhile the minority carrier lifetime in the drift layer is set to be 10 µs, which agrees with the experimental results[28, 29]. The interface state density between SiC and AlN is set to be $10^{12}$ cm$^{-2}$eV$^{-1}$.

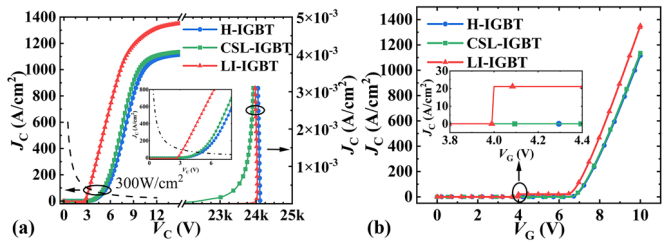

**Figure. 3.** ON-state and OFF-state output characteristics (a) and the transfer characteristics (b) of H-IGBT, CSL-IGBT and LI-IGBT.

The ON-state and OFF-state characteristics of these structures are shown in figure. 3(a). The LI-IGBT shows the same breakdown voltage ($V_{BR}$) of approximately 24 kV compared to that of H-IGBT. The $V_{BR}$ of LI-IGBT is higher than that of CSL-IGBT. Because the CSL increases the leak current for CSL-IGBT and the AlN decreases the leak current by high dielectric constant for LI-IGBT in OFF-state. The collector saturation voltage ($V_{C(sat)}$) (at 100 A/cm$^2$, $V_G$ = 10 V) is separately 5.50 V in H-IGBT, 5.19 V in CSL-IGBT and 3.25 V in LI-IGBT with 50 W/cm$^2$ and 370 nm[30]. The collector staturation current is separately 1115.16 A/cm$^2$ in H-IGBT, 1135.37 A/cm$^2$ in CSL-IGBT, and 1351.41 A/cm$^2$ in LI-IGBT with 50 W/cm$^2$ and 370 nm. Compared to H-IGBT and CSL-IGBT, the saturation current of LI-IGBT has increased by only about 20%. This indicates that while the short-circuit capability of LI-IGBT is still not significantly inferior to those of H-IGBT and CSL-IGBT. Additionally, the short-circuit capability of LI-IGBT can be improved by reducing the driving voltage and increasing the gate resistance[31]. The Baliga's figure of merit (BFOM, $V^2_{BR}/R_{ON\,SP}$)[16] is used to evaluate the trade-off between ON-state and OFF-state characteristics. The BFOM of LI-IGBT is 119.55 GW/cm$^2$ indicating 304.59% and 263.67% improvement compared with CSL-IGBT and H-IGBT.

The transfer characteristics of all structures are illustrated in figure. 3(b) at $V_C$ = 15 V. From figure. 3(b), the collector current density in LI-IGBT shows a sudden increase at $V_G$ = 4 V due to the triggering-on of the integrated LEDs. Then, in the range from $V_G$ = 4 V to $V_G$ = 6.27 V, the collector current density keeps stable. This because the LED is turned on while the MOS-channel is not formed in IGBT. Therefore, the large injection effect does not appear in the drift layer due to the limited photogenerated carriers.

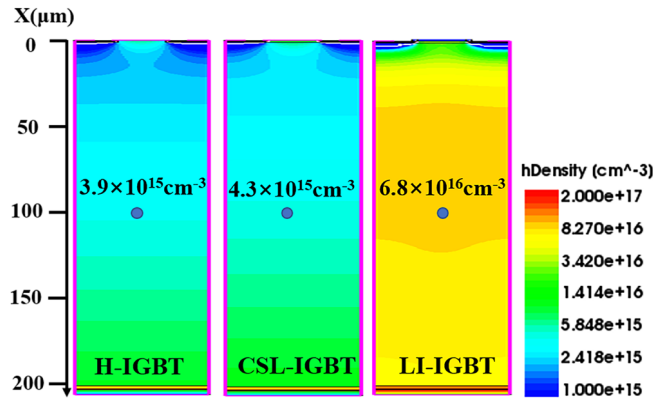

**Figure. 4.** Hole density distribution of H-IGBT, CSL-IGBT and LI-IGBT at a gate bias of 10 V and collector voltage of 5 V.

Figure. 4 shows the hole density distribution of H-IGBT, CSL-IGBT and LI-IGBT (50 W/cm$^2$, 370 nm) at $V_G$ = 10 V and $V_C$ = 5 V. Here, for a given bias condition of the device the distributions of hole and electron is nearly the same. At the middle of drift layer, the hole density of LI-IGBT is 6.8×10$^{16}$ cm$^{-3}$, and it's 17.44 times and 15.81 times higher than the H-IGBT and CSL-IGBT, respectively. This confirms that the UV light can effectively increase the carrier density in the drift layer, indicating high current density at the same collector voltage.

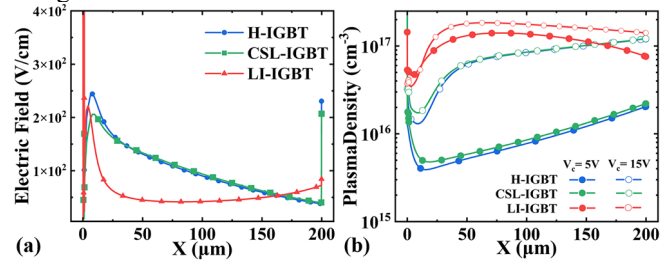

**Figure. 5.** (a) The plasma density and electric field intensity of H-IGBT, CSL-IGBT and LI-IGBT at a gate bias of 10 V and collector voltage of 5 V. (b) The plasma density of H-IGBT, CSL-IGBT and LI-IGBT at a gate bias of 10 V and different collector voltages.

Figure. 5(a) and (b) present the distributions of the electric field and plasma density along the dotted line A in figure.1 for the H-IGBT, CSL-IGBT and LI-IGBT (50 W/cm$^2$, 370 nm) at $V_G$ = 10 V and $V_C$ = 5 V. The distribution behavior of plasma density in the drift layer of LI-IGBT is different from those in CSL-IGBT and H-IGBT. In the middle JFET region, the plasma density of LI-IGBT is 4.76×10$^{16}$ cm$^{-3}$, H-IGBT is 1.18×10$^{16}$ cm$^{-3}$ and CSL-IGBT is 1.76×10$^{16}$ cm$^{-3}$. Meanwhile, at the middle of drift layer in figure. 1, the plasma density of LI-IGBT is 1.37×10$^{17}$ cm$^{-3}$, and is 17.56 times and 15.99 times higher than H-IGBT and CSL-IGBT, respectively. The conventional IGBT structures exhibit a catenary carrier distribution within the SiC epitaxial layer. However, the LI-IGBT has significant advantages in improving carrier concentration in both the JFET and the n-drift regions. The phenomenon is induced by two reasons. First, in the LI-IGBT, injection of carriers from the p+ collector region (bottom side) and the photogenerated carriers from the gate electrode regions (top side) simultaneously produce high-level injection conditions within the n-drift regions. Second, the electric field in the LI-IGBT at the middle of drift layer is almost 2 times lower than that in the H-IGBT and CSL-IGBT. Thus, the injection efficiency is improved in LI-IGBT based upon the electron current continuous equation with the combination of the drift and diffusion parts. In addition, due to the high dielectric constant of the AlN layer, low electric field can be induced in the portion below the gate electrode. Combining it with high density of carriers in the JFET regions, low forward voltage drop occurs in this portion and results in less influence on extraction effect of the minority carriers because the PN junction in the JFET region is reverse biased in the ON-sate.

Figure. 5(b) exhibits the plasma distributions along the dotted line A in figure.1 for H-IGBT and LI-IGBT (50 W/cm$^2$, 370 nm) with different $V_C$ ($V_G$ = 10 V). With increasing collector voltage, the injection efficiency is improved. Thus, the conductivity modulation effect is enhanced with high collector voltage. At the middle of drift layer, the plasma density of the LI-IGBT, H-IGBT and CSL-IGBT increased to 1.29, 9.99 and 10.86 times from $V_C$ = 5 V to $V_C$ = 15 V, respectively. This result demonstrates that, with the integration of 370 nm LED, the plasma concentration in the proposed device shows negligible dependence on $V_C$ due to the fact of the photogenerated carriers for LI-IGBT. The photogenerated carriers with a high-level injection concentration stimulate almost the full potential of conductivity modulation in LI-IGBT and decrease the influence of the large injection of the p+ collector regions.

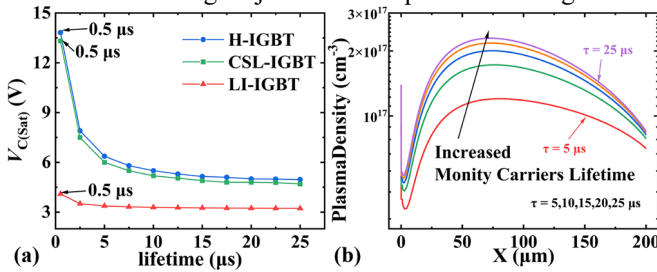

**Figure. 6.** (a) The $V_{C(sat)}$s with different wavelength. (b) The optical generation of LI-IGBT with different wavelength at the gate bias of 10V and the collector bias of 5V.

Figure. 6 shows the $V_{C(sat)}$s and the plasma density with different wavelengths. From figure. 6(a), the $V_{C(sat)}$s in LI-IGBT ($V_G$ = 10 V, 50 W/cm$^2$ LED) exhibit an approximate catenary tendency as the function of wavelength. The minimum $V_{C(sat)}$ appears at 370 nm. There are two reasons for this result. On one hand, the optical generation of the short wavelength light decrease rapidly when light penetrates inside silicon carbide. As shown in figure. 6(b), the optical generation of 280 nm and 340 nm stop at 17.33 μm and 125.4 μm, respectively. On other hand, the energy of long-wavelength light is low, leading to a low rate of stimulated emission for the photogenerated carriers. The optical generation of 370 nm and 380 nm devices are separately with the values of 4×10$^{17}$ cm$^{-3}$ and 1.3×10$^{17}$ cm$^{-3}$. Therefore, the device with 370 nm LED has better ON-state performance than that of 380 nm.

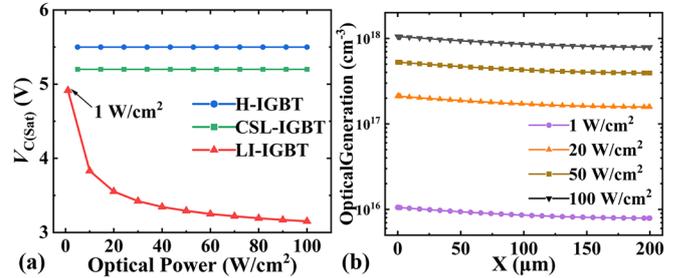

**Figure. 7.** (a) The $V_{C(sat)}$s with different optical power. (b) The optical generation of LI-IGBT with different optical power at the gate bias of 10V and the collector bias of 5V.

The $V_{C(sat)}$s of LI-IGBT with a different optical power ($V_G$ = 10 V, 370 nm LED) is illustrated in figure. 7(a). The $V_{C(sat)}$ is significantly reduced at the optical power of 50 W/cm$^2$ owing to the carriers caused by light inject effect and the improvement of conductivity modulation. Then, the $V_{C(sat)}$ decreases slowly with the further increase of optical power. This can be attributed to the reason that the (Auger) recombination lifetime is decreased at a large density of holes and electrons in the case at high injection levels. When $V_G$ = 10 V and $V_C$ = 5 V, the optical generation along the dotted line A in figure.1 is depicted in figure. 7(b). At the middle of drift layer, the optical generation of 20 W/cm$^2$ device is 20 times higher than that of 1 W/cm$^2$ device, resulting in 27.73% improvement in the $V_{C(sat)}$. Furthermore, the optical generation of 100 W/cm$^2$ device is 2 times of 50 W/cm$^2$. Then the $V_{C(sat)}$ of 100 W/cm$^2$ is 4.28% lower than that of 50 W/cm$^2$.

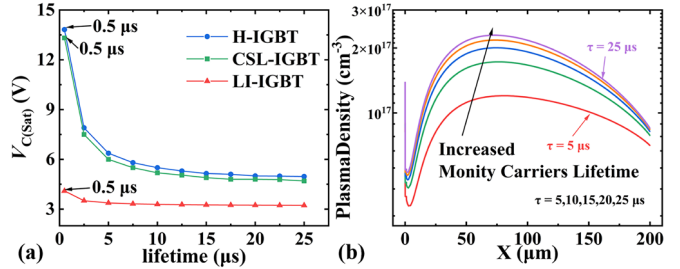

**Figure. 8.** (a) The $V_{C(sat)}$s with different minority carrier lifetime of H-IGBT, CSL-IGBT and LI-IGBT (50 W/cm$^2$, 370 nm). (b) The plasma density of LI-IGBT (50 W/cm$^2$, 370 nm) with different minority carrier lifetime at the gate bias of 10V and the collector bias of 5V.

Figure. 8 compares the effects of the minority carrier lifetimes on the characteristics of the LI-IGBT (50 W/cm$^2$, 370 nm), CSL-IGBT, and H-IGBT. For the conventional devices, long minority lifetime leads to decreased $V_{C(sat)}$ due to low recombination rate of carriers. With respect to the LI-IGBT, the dependence of LI-IGBT decreases with the increase of minority carrier lifetime. Such phenomenon implies that the improvement in ON-state performance of the LI-IGBT does not only rely on the increment of the minority carrier lifetime can be realized, but also significantly rely on the wavelength and optical power of the integrated LED (figures. 6 and 7). It shows that light-emitting-diodes integrated in the gate is a new way to improve the electric characteristics and reduce the impact of the high density of deep-level defects in SiC IGBT. It can be seen that, with the lifetime of 0.5 μs and 25 μs, the LI-IGBT, CSL-IGBT, and H-IGBT separately have 21.21%, 64.03%, and 64.80% improvement in $V_{C(sat)}$. For the low minority carrier lifetime in LI-IGBT, there are still large amounts of photogenerated carriers to enhance the conductivity modulation in the drift layer and the JFET region. Then, $V_{C(sat)}$ remains almost constant with the further increase of the carrier lifetime. Because large amounts of the photogenerated carriers manifest almost all the region of drift layer. Therefore, the carriers do not need long time to drift or diffuse in the drift layer. At $V_G = 10$ V and $V_C = 5$ V, the plasma density along the dotted line A of LI-IGBT (50 W/cm$^2$, 370 nm) with a different lifetime is illustrated in figure. 8(b). At the middle of drift layer, the plasma density of LI-IGBT increases from $9.60\times10^{16}$ cm$^{-3}$ to $1.80\times10^{17}$ cm$^{-3}$ when the minority carrier lifetime increases from 5 μs to 25 μs.

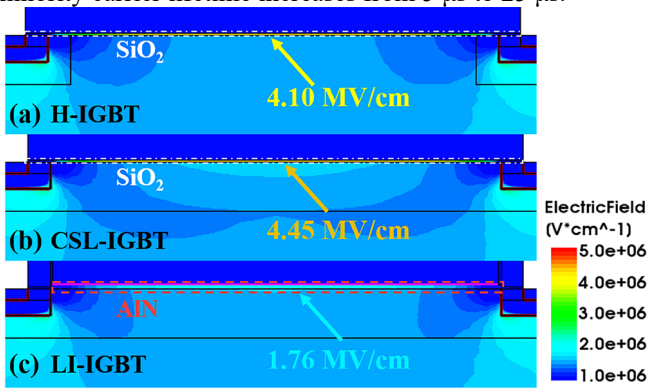

**Figure. 9.** The OFF-state electric field in the (a) H-IGBT, (b) CSL-IGBT and (c) LI-IGBT.

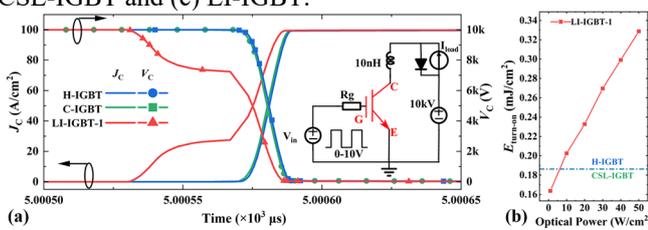

**Figure. 10.** (a) The turn-on waveform of H-IGBT, CSL-IGBT and LI-IGBT-1 (50 W/cm$^2$, 370 nm), the inset shows the circuit schematic of dynamic analysis. (b) The turn-on loss of LI-IGBT-1 with different optical power.

Figure. 9 illustrates the electric field distributions of H-IGBT, CSL-IGBT and LI-IGBT at OFF-state ($V_G = 0$ V, $V_C = 24$ kV). The electric field of the dielectric layer in LI-IGBT (1.76 MV/cm) is approximately half of those in H-IGBT and CSL-IGBT. Such phenomenon can be attributed to two primary factors. First, the dielectric coefficient of AlN is approximately 7.5, whereas that of SiO$_2$ is 3.9. Consequently, under the same OFF-state collector voltage, the electric field in the LI-IGBT is lower than those in H-IGBT and CSL-IGBT as explained by Gauss's law. Second, the grounded cathode of the integrated LED further contributes to the reduction of the electric field under the gate area in the LI-IGBT.

Figure. 10(a) illustrates the comparisons of the turn-on transients of H-IGBT, CSL-IGBT and LI-IGBT-1 (50 W/cm$^2$, 370 nm) and the inset shows the circuit schematic of dynamic analysis. The parasitic inductance is set to be 10 nH. The simulations are carried out at a bus voltage of 10 kV and current of 100 A/cm$^2$, respectively. The operation frequency is 200 Hz[20]. The turn-on time and the turn-on loss of LI-IGBT-1 are 0.045 μs and 71.79 mJ/cm$^2$, respectively. All of them are larger than those of H-IGBT (0.010 μs, 18.65 mJ/cm$^2$) and CSL-IGBT (0.011 μs, 18.60 mJ/cm$^2$). The LI-IGBT-1 turns on at low current density via the active LED that has the threshold voltage of 4 V. It is noteworthy that this device generates greater loss than H-IGBT and CSL-IGBT. Figure. 10(b) shows the turn-on loss of LI-IGBT-1 with different optical power. The lowest turn-on loss is observed at an optical power of 1 W/cm$^2$, with the turn-on time of 0.008 us. This performance can be attributed to the LI-IGBT operating at a wavelength of 370 nm, where the application of 1 W/cm$^2$ of optical power effectively activates the device to the ON-state through a moderate generation of photogenerated carriers.

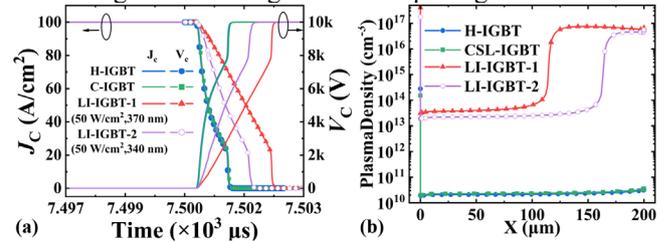

**Figure. 11.** (a) The turn-off waveform of H-IGBT, CSL-IGBT and LI-IGBT-1 (50 W/cm$^2$, 370 nm), LI-IGBT-2 (50 W/cm$^2$, 340 nm). (b) The plasma density in the drift layer of H-IGBT, CSL-IGBT and LI-IGBT-1, LI-IGBT-2, after the gate is turn off for 0.3 μs.

Comparisons of the turn-off waveforms are shown in figure. 11(a). Here, due to the area factor of LED is approximately one ten thousandth of LI-IGBT, the loss of integrated LED is roughly three orders of magnitude less than that of IGBT. Thus, the loss of LED can be neglected. From figure. 11(a), The turn-off time and the turn-off loss of LI-IGBT-1 are 1.63 μs and 328.71 mJ/cm$^2$, respectively. They are 1.50 μs and 232.75 mJ/cm$^2$ in LI-IGBT-2. All of them are larger than those of H-IGBT (0.73 μs, 161.42 mJ/cm$^2$) and CSL-IGBT (0.73 μs, 159.69 mJ/cm$^2$). Figure. 11(b) shows the turn-off plasma density of the devices along the dotted line A in figure.1. After the gate is turned off for 0.3 μs, the plasma density of H-IGBT and CSL-IGBT are $\sim3\times10^{10}$ cm$^{-3}$, but the plasma density of LI-IGBTs are all above $1\times10^{13}$ cm$^{-3}$. This is because amounts of photogenerated carriers in LI-IGBT have put a burden on the recombination of carriers in turn-off transient. Additionally, compared to the LI-IGBT-2, LI-IGBT-1 has a higher turn-off loss owing to more photogenerated carriers induced by the sufficient optical power and the appropriate wavelength.

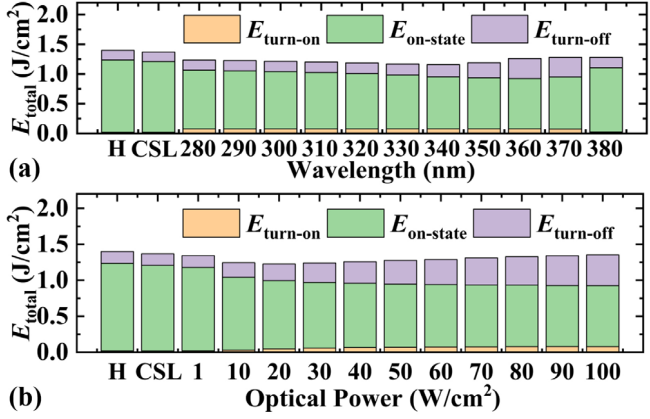

**Figure. 12.** (a) The energy loss of H-IGBT, CSL-IGBT and LI-IGBT (50 W/cm$^2$) with different wavelength. (b) The loss of H-IGBT, CSL-IGBT and LI-IGBT (370 nm) with different optical power.

Inspiringly, for the loss of LI-IGBT, it can be decreased by selecting LED with the matched optical power and wavelength. The energy loss consists of turn on loss, ON-state loss and turn off loss. The energy losses of the LI-IGBT with different wavelength and optical power are shown in figure. 12 (a) and (b), respectively. From these results it can be seen that the LI-IGBT exhibits less energy loss than H-IGBT and CSL-IGBT because of a significant improvement in forward characteristics. The minimum energy loss of LI-IGBT with 50 W/cm$^2$ LED is 1.15 J/cm$^2$ at 340 nm, whereas the minimum energy loss of LI-IGBT with 370 nm LED is 1.18 J/cm$^2$ at 20 W/cm$^2$. These values are approximately 14%-18% lower than those of H-IGBT and CSL-IGBT. Therefore, a superior trade-off between $V_{C(sat)}$ and total energy loss can be attained in the proposed structure by adjusting the optical power and wavelength of the LED.

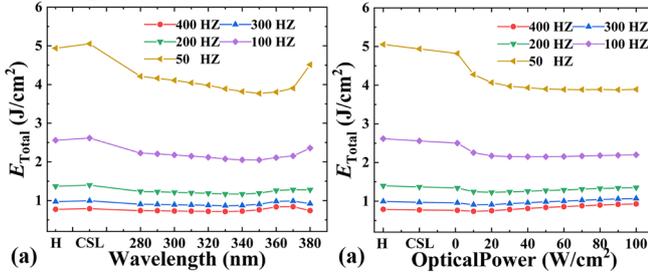

**Figure. 13.** (a) The energy loss of H-IGBT, CSL-IGBT and LI-IGBT (50 W/cm$^2$) with different wavelength at different frequencies. (b) The loss of H-IGBT, CSL-IGBT and LI-IGBT (370 nm) with different optical power at different frequencies.

The energy losses of the LI-IGBT with different wavelength are shown in figure. 13 (a). With a duty cycle $\delta$, $E_{total}$ involves in the ON-state power loss and the switching power dissipation [16]

$$E_{\text{total}} = \frac{ft_{ON}}{\delta} f R_{ON \cdot SP} I_{ON}^2 + \frac{\delta - ft_{ON}}{2\delta} I_{ON} V_{CE} \quad (1)$$

Where $\delta$ represents the ratio of the ON-state interval to the switching cycle, f is the switching frequency, $t_{ON}$ is the turn-on time of the IGBT, $I_{ON}$ is the ON-state current of the IGBT,

$R_{ON \cdot SP}$ is the resistance of the IGBT at the $I_{ON}$, and $V_{CE}$ is the collector voltage of the IGBT. As shown in formula (1), the proportion of energy losses in the ON-state increases as the frequency decreases. Therefore, when the frequency is below 400 Hz, the total loss of LI-IGBT (50 W/cm$^2$) at different wavelengths is less than that of H-IGBT and CSL-IGBT. Furthermore, when the frequency is at 400 Hz, the total loss of LI-IGBT (50 W/cm$^2$) from 280 nm to 340 nm is less than that of H-IGBT and CSL-IGBT due to the superior balance of ON-state losses and switch losses. The energy losses of the LI-IGBT with different optical power are shown in figure. 13(b). Likewise, when the frequency is below 300 Hz, the total loss of LI-IGBT (370 nm) with different optical powers is less than that of H-IGBT and CSL-IGBT. However, when the frequency is at or above 300 Hz, the total loss of 10 W/cm$^2$ LI-IGBT (370 nm) is lower than that of H-IGBT and CSL-IGBT.

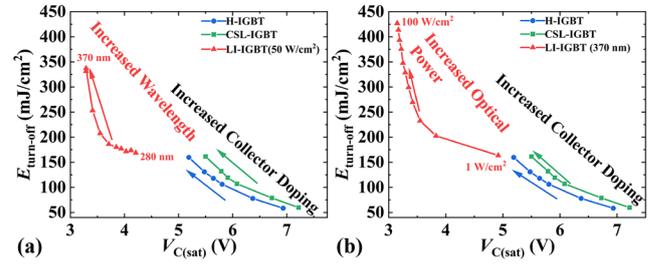

**Figure. 14.** (a) The trade-off relationship between $V_{C(sat)}$ and $E_{turn-off}$ in H-IGBT and CSL-IGBT with different collector doping, and that in LI-IGBT (50 W/cm$^2$) with different wavelength. (b) The trade-off relationship between $V_{C(sat)}$ and $E_{turn-off}$ in H-IGBT and CSL-IGBT with different collector doping, and that in LI-IGBT (370 nm) with different optical power.

Figure. 14(a) shows the trade-off plots of $V_{C(sat)}$ versus $E_{turn-off}$ in H-IGBT and CSL-IGBT with different collector doping, and those in LI-IGBT (50 W/cm$^2$) with the wavelength from 280 nm to 370 nm. The collector doping levels for H-IGBT and CSL-IGBT have been increased from $5 \times 10^{18}$ cm$^{-3}$ to $1 \times 10^{19}$ cm$^{-3}$. To achieve the lowest $V_{C(sat)}$, the value is 5.50 V for H-IGBT with $E_{turn-off}$ of 161.42 mJ/cm², and is 5.19 V for CSL-IGBT with $E_{turn-off}$ of 159.69 mJ/cm². In the case of LI-IGBT with an optical power of 50 W/cm², the optimal trade-off is attained at 280 nm, resulting in $V_{C(sat)}$ of 4.21 V and $E_{turn-off}$ of 168.87 mJ/cm². Figure. 14(b) illustrates the trade-off relationship of $V_{C(sat)}$ and $E_{turn-off}$ in H-IGBT and CSL-IGBT with different collector doping, and those in LI-IGBT (370 nm) with different optical power. For LI-IGBT at a wavelength of 370 nm, the optimal trade-off occurs at 1 W/cm², resulting in $V_{C(sat)}$ of 4.92 V and $E_{turn-off}$ of 163.7 mJ/cm². This is attributed to the restricted number of photo-generated carriers under these conditions, in contrast to other scenarios. For H-IGBT and CSL-IGBT, the lowest $V_{C(sat)}$ occurs at collector doping of $1 \times 10^{19}$ cm$^{-3}$. Therefore, a superior trade-off between $V_{C(sat)}$ and $E_{turn-off}$ can be realized by selecting appropriate LED wavelength and optical power.

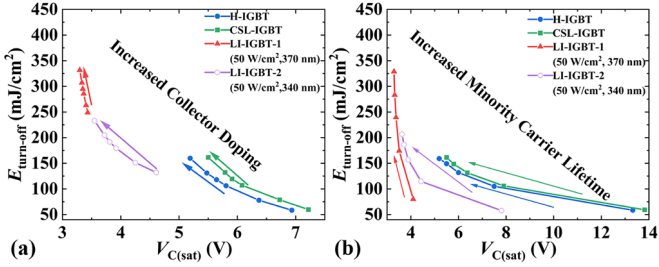

**Figure. 15.** The trade-off relationship between $V_{C(sat)}$ and $E_{turn\text{-}off}$ in H-IGBT and CSL-IGBT, in LI-IGBT-1 (50 W/cm², 370 nm), and in LI-IGBT-2 (50 W/cm², 340 nm), with (a) different collector doping and (b) different minority carrier lifetime.

Figure. 15(a) illustrates the trade-off relationship of $V_{C(sat)}$ and $E_{turn\text{-}off}$ in H-IGBT, CSL-IGBT, and LI-IGBT-1 (50 W/cm², 370 nm), LI-IGBT-2 (50 W/cm², 340 nm), with different collector doping, respectively. The collector doping has increased from $5 \times 10^{18}$ cm$^{-3}$ to $1 \times 10^{19}$ cm$^{-3}$. For LI-IGBT-1, the superior adaptability of SiC to 370 nm light absorption and penetration depth results in negligible alterations to its trade-off curves of $V_{C(sat)}$ and $E_{turn\text{-}off}$. However, for LI-IGBT-2 with LED of 50 W/cm², 340 nm, the optimal trade-off is observed at a collector doping of $5 \times 10^{18}$ cm$^{-3}$, which results in $V_{C(sat)}$ of 4.61 V and $E_{turn\text{-}off}$ of 132.03 mJ/cm². For comparison, $E_{turn\text{-}off}$ is 161.42 mJ/cm² in H-IGBT with $V_{C(sat)}$ of 5.50 V and is 159.69 mJ/cm² in CSL-IGBT with $V_{C(sat)}$ of 5.19 V. This enhancement of turn-off performance of LI-IGBT-2 is attributed to the reduction in collector injection efficiency. Figure. 15(b) illustrates the trade-off relationship of $V_{C(sat)}$ and $E_{turn\text{-}off}$ in LI-IGBT-1 (50 W/cm², 370 nm), LI-IGBT-2 (50 W/cm², 340 nm), H-IGBT and CSL-IGBT with different minority carrier lifetime. The minority carrier lifetime has increased from 0.5 µs to 10 µs. The superior trade-off is shown in LI-IGBT-1 when the minority carrier lifetime is 0.5 µs, which results in $V_{C(sat)}$ of 4.10 V and $E_{turn\text{-}off}$ of 80.03 mJ/cm². In LI-IGBT-2, $E_{turn\text{-}off}$ is 58.55 mJ/cm² with and $V_{C(sat)}$ is 7.81 V. For comparison, $E_{turn\text{-}off}$ is 58.55 mJ/cm² in H-IGBT with $V_{C(sat)}$ of 13.32 V and is 59.33 mJ/cm² in CSL-IGBT with $V_{C(sat)}$ of 13.82 V. This is because large amounts of photogenerated carriers enhance the conductivity modulation in the drift layer and the JFET region at ON-state. However, at OFF-state, the low minority carrier lifetime facilitates rapid carrier recombination. Thus, the trade-off of LI-IGBT can be effectively optimized by reducing collector doping level and minority carrier lifetime.

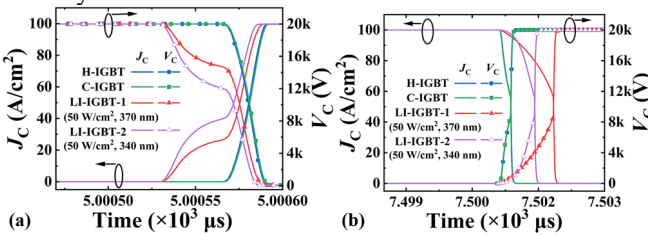

**Figure. 16.** The (a) turn-on waveform and (b) turn-off waveform of H-IGBT, CSL-IGBT and LI-IGBT-1 (50 W/cm², 370 nm), LI-IGBT-2 (50 W/cm², 340 nm) at 20 kV.

Figure. 16(a) illustrates the turn-on transients of H-IGBT, CSL-IGBT and LI-IGBT-1 (50 W/cm², 370 nm), LI-IGBT-2 (50 W/cm², 340 nm) at 20 kV. The performance characteristics at a bus voltage of 20 kV are similar to those at 10 kV. The turn-on times for LI-IGBT-1 and LI-IGBT-2 are 0.047 µs and 0.046 µs, respectively, with turn-on losses of 15.99 mJ/cm² and 18.46 mJ/cm². In comparison, the turn-on times for H-IGBT and CSL-IGBT are 0.015 µs and 0.011 µs, respectively, with turn-on losses of 5.85 mJ/cm² and 5.71 mJ/cm². Figure. 16(b) illustrates the turn-off transients of the four devices at 20 kV. The turn-off times for LI-IGBT-1 and LI-IGBT-2 are 0.75 µs and 0.36 µs, respectively, resulting in shutdown losses of 304.18 mJ/cm² and 130.94 mJ/cm². For H-IGBT and CSL-IGBT, the turn-off times are 0.21 µs and 0.23 µs, with turn-off losses of 87.7 mJ/cm² and 88.59 mJ/cm², respectively. At a bus voltage of 20 kV, the turn-off times and losses for the devices are reduced compared to a 10 kV bus voltage. This improvement occurs because the higher bus voltage facilitates a quicker extraction of charge carriers during turn-off transient.

**Table 2.** Performance comparisons between LI-IGBT, H-IGBT, and CSL-IGBT

| Parameters | LI-IGBT (50 W/cm², 370 nm) | H-IGBT | CSL-IGBT |
|---|---|---|---|
| $V_B$ (kV) | 24.03 | 24.03 | 23.95 |
| $R_{ON\cdot SP}$ (mΩ·cm²) | **4.83** | 14.71 | 12.65 |
| $V_{C(sat)}$ (V) @ $J_C$ = 100 A/cm² | **3.25** | 5.50 | 5.19 |
| $E_{total}$ (J/cm²) | **1.28** | 1.37 | 1.40 |
| BFOM = $V^2_{BR}/R_{ON,SP}$ (GW/cm²) | **119.55** | 39.25 | 45.34 |

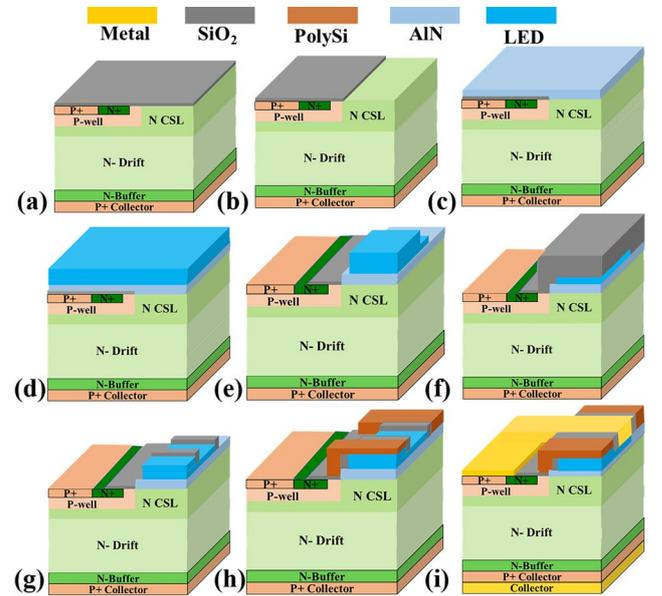

**Figure 17.** Process flow of the growth of integrated LEDs of SiC LI-IGBT. (a) Oxide growth. (b) Etching the oxide. (c) Deposition of AlN on top of CSL. (d) Growth of integrated LED. (e) Etching a part of LED and AlN. (f) Deposition of oxide. (g) Etching the oxide. (h) Deposition of polysi. (i) Formation of emitter.

Eventually, the static and dynamic characteristics of the compared SiC IGBT structures are summarized in Table 2. Among all the devices, the LI-IGBT exhibits comparable breakdown voltage to H-IGBT and CSL-IGBT. Furthermore, it features both lower $R_{ON \cdot SP}$ (4.83 mΩ·cm$^2$) and $V_{C(sat)}$ (3.25 V) compared to other devices due to the improvement of the carriers density in the JFET region and the drift layer. Additionally, Baliga's figure of merit (BFOM) of the LI-IGBT shows the highest value of 119.55 GW/cm$^2$ in comparison to H-IGBT and CSL-IGBT, enabling the considerable enhancements in both the ON-state and Off-state performances for SiC IGBTs.

Figure 17 shows the cross-sectional process flow diagrams of the integration of LEDs into SiC LI-IGBT. After all the ion implantation doping of the IGBT, the oxide is grown by thermal oxide (figure. 17(a)). Then, the oxide on the JFET region is etched by inductively coupled plasma (ICP)[32] (figure. 17(b)). The AlN is deposited as a buffer layer via metal-organic chemical vapor deposition (MOCVD)[25] (figure. 17(c)) to protect the oxide on the channel. The microstructure of LED is formed by epitaxial growth (figure. 17(d)). Then, the reactive ion etching (RIE) is used to expose the oxide surface of SiC apart from the channel regions. Thereafter, the ICP is used to expose the contact surface of the n/p regions in LED (figure. 17(e)). For the isolation of the p contact and n contact of LED, the low temperature atomic layer deposition (ALD) is used to deposit the oxide (figure. 17(f-g))[33]. The low-pressure chemical vapor deposition (LPCVD) is used to deposit the polysi (figure. 17(h))[34]. The metal is deposited as the emitter of IGBT and the cathode electrode of LED (figure. 17(i)). The presentation of the fabrication flow verifies the fact that the integration process of LED is compatible with manufacturing SiC power device.

## 4. Conclusion

In conclusion, a novel 4H-SiC LI-IGBT is proposed and demonstrated via TCAD simulations. Unlike the traditional IGBT, the static and dynamic performances of the LI-IGBT can be controlled by the wavelength and optical power of the integrated LEDs. The photogenerated carriers stimulated by the LEDs can significantly improve the plasma density in the drift region of IGBT, resulting in a low forward voltage drop in it. The LI-IGBT with embedded 50 W/cm$^2$, 370 nm LED shows the superior performance compared to other devices. The forward voltage drop of LI-IGBT is reduced by 40.90% and 37.38% compared with H-IGBT and CSL-IGBT, respectively. The BFOM of LI-IGBT is 119.55 MW/cm$^2$, which is separately 304.59% and 263.67% higher than those of the CSL-IGBT and H-IGBT. In addition, the loss of LI-IGBT is separately reduced by 6.57% and 8.57% compared to H-IGBT and CSL-IGBT in one cycle. Meanwhile, the relationship between $V_{C(sat)}$ and $E_{turn-off}$ can be optimized by adjusting collector doping level and minority carrier lifetime. These results demonstrate that the SiC LI-IGBT shows high degree of accessibility and can realize fruitful targets in ultra-high voltage application.

## Acknowledgements

This research was supported by the National Natural Science Foundation of China (Grant No. 62274137), Natural Science Foundation of Jiangxi Province of China for Distinguished Young Scholars (No. S2021QNZD2L0013), the Fundamental Research Funds for the Central Universities (Grant No. 20720230103), National Key Research and Development Program of China (Grant No. 2023YFB3609500).